# Phase-preserving chirped-pulse optical parametric amplification to 17.3 fs directly from a Ti:Sapphire oscillator

C. P. Hauri, P. Schlup, G. Arisholm[1],

J. Biegert, U. Keller

*Swiss Federal Institute of Technology (ETH Zürich), Physics Department,*

*CH-8093 Zürich, Switzerland*

*Phone: +41 1 633 65 30, fax: + 41 1 633 10 59, email: biegert@phys.ethz.ch*

Phase-stabilized 12-fs, 1-nJ pulses from a commercial Ti:Sapphire oscillator are directly amplified in a chirped pulse optical parametric amplifier (CPOPA) and recompressed to yield near-transform-limited 17.3-fs pulses. The amplification process is demonstrated to be phase preserving and leads to 85-µJ, carrier-envelope-offset (CEO) phase-locked pulses at 1 kHz for 0.9 mJ of pump, corresponding to a single-pass gain of $8.5 \times 10^4$. © 2004 Optical Society of America

Chirped-pulse optical parametric amplification (CPOPA) [1-3] is rapidly emerging as an attractive alternative to conventional stimulated emission-based chirped pulse amplifier (CPA) systems for the amplification of ultrashort pulses. Large single-pass parametric gains, on the order of $10^7$, are in principle possible by propagation through only millimeters of material, yielding substantially reduced *B* integrals; the gain

---

[1] Current address: Forsvarets forskningsinstitutt (Norwegian Denfence Research Establishment), P.O. Box 25, NO-2027 Kjeller, Norway

bandwidth can be tailored by choice of nonlinear optical crystal and interaction geometry, with bandwidths in excess of 180 THz (6000 cm$^{-1}$) previously reported [4]; and since only transitions between virtual states are involved, there is no energy storage and thermal loading is virtually eliminated, which is advantageous for high repetition rate applications. The major limitation to CPOPA has been the availability of pump sources capable of delivering sufficiently short, high-energy pulses. Even so, with existing technology two extreme features have been demonstrated: multi-terawatt-level amplification with a long pulse, low-repetition rate, Nd:glass laser [5,6], and ultrabroadband amplification to yield sub-5-fs pulses at the few-microjoule level from a white-light seed [4,7,8]. Here, we choose a picosecond pump source since it represents an ideal compromise between pulses short enough to allow for bulk stretching and prism compression of the seed [9], avoiding the potentially phase-disturbing influences of diffraction gratings [10]; but sufficiently long to alleviate the need for precise pulse-front matching, necessary with femtosecond pump pulses to permit accurate recompression without spatial chirp [8]. In theory, the phase of the amplified seed remains, aside from quantum noise, unaltered by amplification with a non-stabilized pump since the idler field dissipates the phase offset. This makes CPOPA eminently suitable for applications such as high-harmonic generation with few-cycle pulses in which the carrier–envelope offset (CEO) phase [11,12] is of paramount importance. Experimental verification of the phase preservation in CPOPA has, to our knowledge, not been previously reported.

In this letter, we demonstrate the phase preservation of CPOPA by directly amplifying the output of a phase-stabilized oscillator, using the experimental configuration schematically illustrated in Fig. 1. Traditional short-pulse OPA/CPOPA systems are seeded by white-light continua at the so-called "magic" visible-wavelength broadband phase matching angle in BBO [13], but would require arduous phase-stabilization schemes for the white-light pump to benefit from the phase preservation in the CPOPA process. Direct amplification of Ti:Sapphire oscillator pulses has previously been demonstrated with pumping by nanosecond pulses, and recompression to 310 fs has been reported [14].

As a seed laser, we used a CEO-phase-stabilized (Menlo Systems), commercial Ti:Sapphire oscillator (Femtolasers), that delivered 700 mW of 12-fs pulses at a repetition rate of 76 MHz. The oscillator phase stabilization feedback system required 175 mW of the output power; and 350 mW were directed to the regenerative amplifier. Seed pulses were selected at a 1 kHz repetition frequency, stretched in a DAZZLER (FastLite), amplified in the CPOPA, and recompressed in a prism compressor.

The CPOPA was pumped by the frequency-doubled output from a modified commercial Ti:Sapphire regenerative amplifier system (Spitfire, Positive Light). Seeding the Spitfire with part of the output from the oscillator ensured synchronization between pump and seed pulses in the CPOPA. The Spitfire produced 2.5-mJ pulses with a duration of 4.3 ps (FWHM). The pulses were frequency-doubled in a 2-mm thick BBO crystal, cut for type-I second-harmonic generation (SHG) at 800 nm, with a conversion efficiency of 40%. The resulting 400-nm pump pulses were focused by a 1-m radius of curvature focusing mirror into the CPOPA crystal (3-mm-long BBO crystal cut at $\theta = 29.2°$) to yield a pump intensity of 65 GW/cm$^2$.

The DAZZLER was used as the pulse stretcher for the CPOPA seed, and simultaneously allowed for higher-order dispersion correction during the optimization of the pulse compression. The stretched 1-nJ seed pulses were loosely focused into the CPOPA crystal using a 1-m focal length lens. The pump and seed beams were overlapped in the CPOPA crystal with a noncollinear angle of $\alpha = 2.1°$ for near-degenerate OPA, for which the gain bandwidth exceeded 70 THz. For a pump energy of 0.9 mJ, the 1-nJ seed was amplified to 85 µJ, corresponding to a single pass gain of $8.5\times10^4$. The solid line in Fig. 2(a) shows the measured amplified spectrum, which supported a theoretical transform-limited pulse duration of 17.2 fs. After amplification, the pulses are recompressed in a double-prism compressor, designed using numerical and ray-tracing simulations [9]. The compressor exhibited some 10% transmission losses, reducing the energy available in the compressed pulses to 77 µJ. The compressed pulses were characterized by SPIDER [15,16], and the reconstructed spectral phase variations were minimized using the incident phase adjustment provided by the DAZZLER. The optimized phase is shown by the dashed line in Fig. 2(a); it exhibited phase variations of less than $\pm\pi/4$ across the whole spectral range. The reconstructed, near-transform-limited (17.3±0.2)-fs temporal pulse shape for the optimized pulse is shown in Fig. 2(b). The transverse intensity profile was recorded with a high resolution CCD (DataRay) and is shown by the inset in Fig. 2(b). We anticipate that the CPOPA output will soon be suitable even for high-field physics experiments.

In order to verify the phase preservation in our configuration of CPOPA, we measured the beat signal between the high frequency part of a white-light spectrum, generated in a 1-mm thick sapphire plate, and the low frequency components frequency-doubled in a 250-µm thick BBO crystal [17]. The CEO phase of the amplified pulses could be derived from the spectral location of the interference fringes, which were recorded with either of two spectrometers equipped with linear CCD arrays (USB2000, Ocean Optics; SpectraPro 300i, Acton Research). Figure 3(a) shows the temporal evolution of interference spectra recorded over 15,000 consecutive pulses. With the phase stabilization to the oscillator switched off (Fig. 3(a), top), averaging over 30 pulses due to the integration time of the CCD array smears out the interference since successive pulses have random relative CEO phases, and no fringes are visible. By contrast, the interference fringes for the CEO-phase-stabilized oscillator (bottom) are resolved and stationary, apart from fluctuations introduced by air currents and mechanical vibrations. To our knowledge, this is the first experimental verification of CEO phase preservation in a CPOPA process. The CEO interference fringes for the phase-stabilized CPOPA were observed to remain resolved and stationary over 10,000 pulses, as shown by the solid line in the time-integrated plot of Fig. 3(b), whereas after integration over the same time period for the unstabilized oscillator, shown by the dashed line, manifested no resolved fringes.

In conclusion, we have demonstrated direct amplification of 1-nJ phase-stabilized oscillator pulses to a pulse energy of 85 μJ for a 0.9-mJ pump, and recompressed the amplified pulses to a near-transform limited pulse duration of 17.3 using the DAZZLER. The output energy could be increased directly by multi-passing the CPOPA crystal [8], while numerical modeling predicts further energy scaling with larger beam sizes. Measurements of the CEO phase of the amplified pulses demonstrates that the CPOPA process preserves the seed phase.

The authors wish to express their gratitude to Positive Light for the loan of the Spitfire amplifier, and thank W. Kornelis and F. Helbing for the SPIDER and CEO measurements.

This work was supported by the ETH Zürich and by the Swiss National Science Foundation.

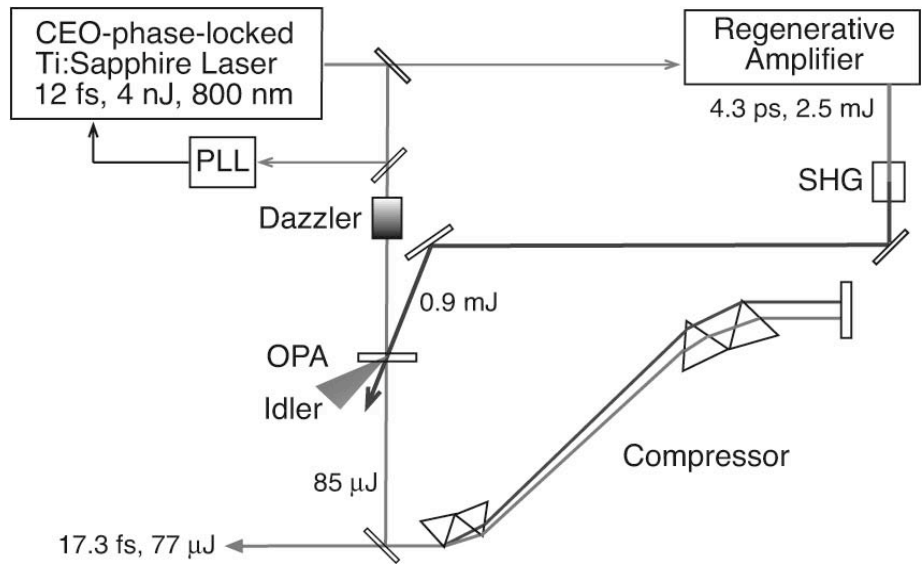

Fig. 1 Experimental configuration: PLL, phase-locked loop to phase-stabilize the oscillator; SHG, second-harmonic generation crystal; Dazzler, bulk stretching and spectral phase adjustment; OPA, 3-mm BBO for near-degenerate phase matched CPOPA.

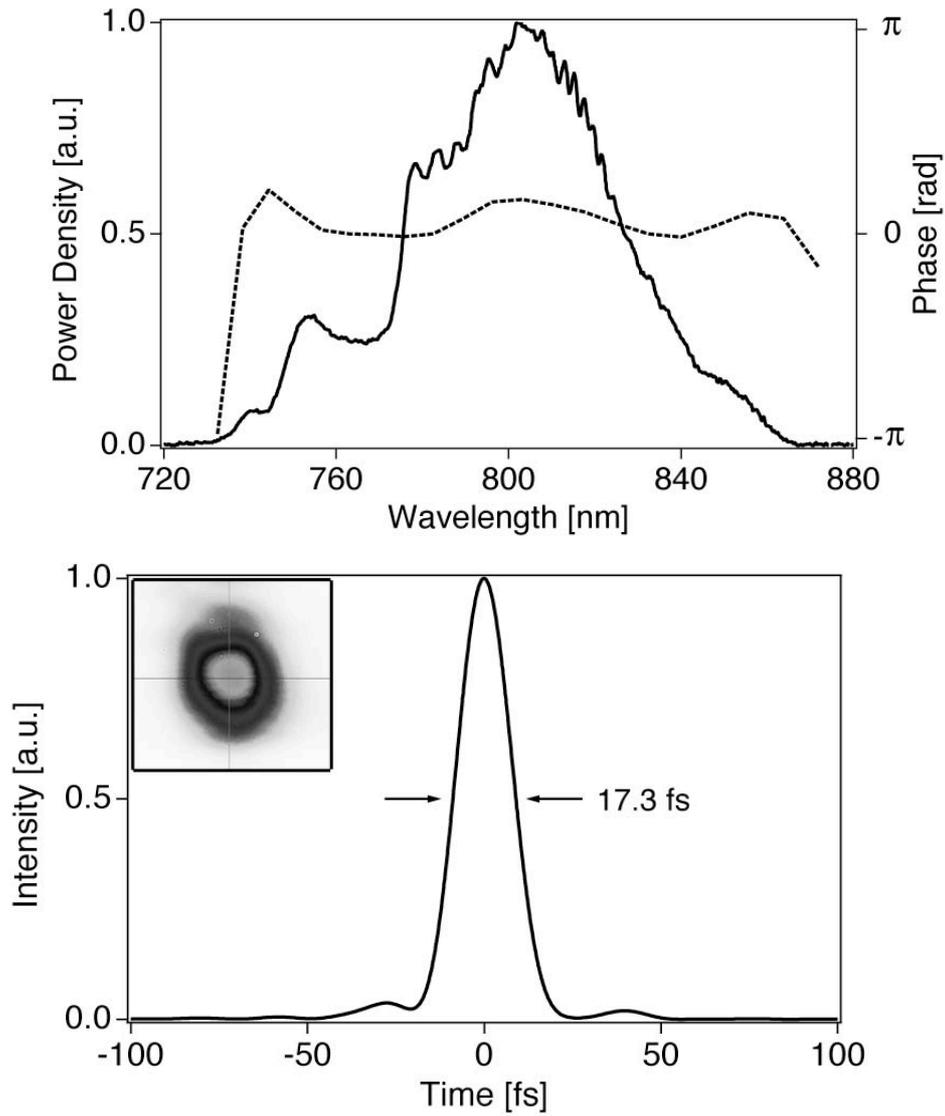

Fig. 2 (a) Amplified pulse spectrum (solid line) and optimized spectral phase (dashed) of the compressed, amplified pulses, measured by SPIDER. (b) Reconstructed pulse profile and (inset) measured far-field spatial intensity distribution.

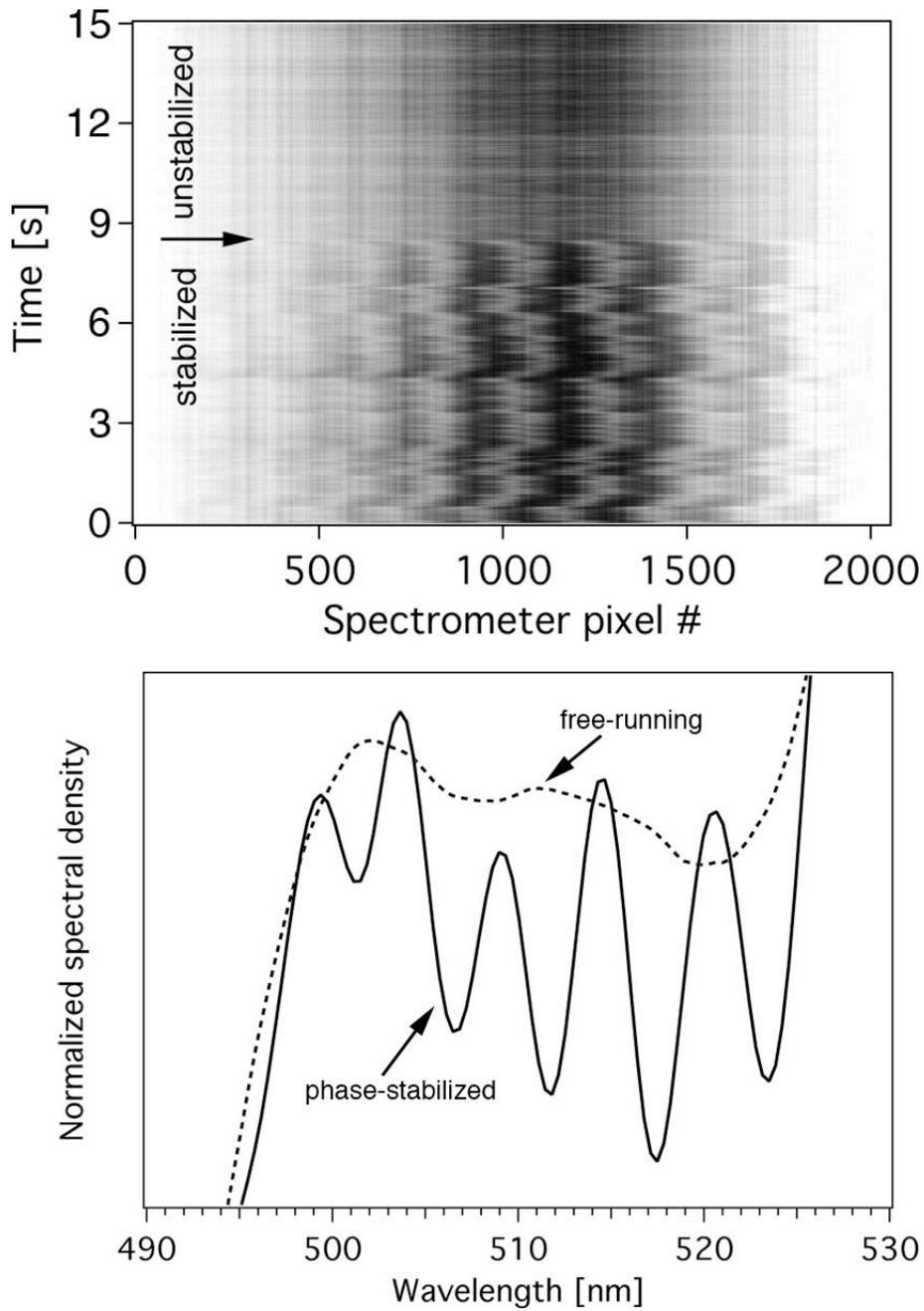

Fig. 3 (a) Temporal evolution of CEO phase measurement interference fringes of the amplified, compressed pulses with (top) free-running and (bottom) phase-stabilized seed pulses. Integration over 5 pulses by the CCD camera blurred the interference fringes when the pulses were not stabilized. (b) Interference fringes averaged over 10,000 shots for free-running (dashed line) and phase-stabilized (solid) seed pulses.